# Data Protection and Corporate Reputation Management in the Digital Era

Gabriela Wojak[1], Ernest Górka[2], Michał Ćwiąkała[3], Dariusz Baran[4], Dariusz Reśko[5], Monika Wyrzykowska-Antkiewicz[6], Robert Marczuk[7], Marcin Agaciński[8], Daniel Zawadzki[9], Jan Piwnik[10]

***Abstract:***

***Purpose:*** *The purpose of this paper is to analyze the relationship between cybersecurity management, data protection, and corporate reputation in the context of digital transformation. The study aims to identify how organizations implement strategies and tools to mitigate cyber risks, comply with regulatory requirements, and build stakeholder trust.*
***Design/methodology/approach:*** *A quantitative research design was applied using an online diagnostic survey distributed to enterprises across various industries in Poland. The study examined organizational cybersecurity strategies, the use of technical and procedural safeguards, employee awareness, and the adoption of international standards such as ISO/IEC 27001 and ISO/IEC 27032.*
***Findings:*** *The research revealed that most organizations have formal cybersecurity strategies, conduct regular audits, and invest in awareness programs. However, despite high levels of preparedness, 75% of surveyed firms experienced cyber incidents, most often resulting in reputational damage and operational disruptions. Cybersecurity is increasingly perceived as an investment in long-term organizational stability rather than as a cost.*

[1]*I'M BRAND INSTITUTE Sp. z o.o., ORCID: 0009-0003-2958-365X, e-mail: g.wojak@imbrandinstitute.com;*
[2]*Department of Social Sciences and Computer Science Nowy Sącz School of Business - National Louis University, Poland, e-mail: ewgorka@wsb-nlu.edu.pl;*
[3]*University College of Professional Education in Wroclaw, ORCID: 0000-0001-9706-864X, e-mail: michal.cwiakala@wskz.pl;*
[4]*The same as in 2, ORCID: 0009-0006-8697-5459, e-mail: dkbaran@wsb-nlu.edu.pl;*
[5]*The same as in 2, ORCID: 0000-0003-4129-0502, e-mail: dresko@wsb-nlu.edu.pl;*
[6]*WSB Merito University in Toruń, Poland, ORCID: 0000-0002-2755-9352 e-mail: monika.wyrzykowska@torun.merito.pl;*
[7]*The same as in 1, ORCID:0009-0008-3553-6581, email: r.marczuk@imbrandinstitute.com;*
[8]*Pomeranian Higher School in Starogard Gdanski, Institute of Management, Economics and Logistics, ORCID: 0009-0009-6003-2913, e-mail: marcin.agacinski@twojestudia.pl;*
[9]*GLOBAL HYDROGEN spółka akcyjna Poland, ORCID: 0009-0001-4783-3240, e-mail: daniel.zawadzki@globalhydrogen.pl;*
[10]*WSB Merito University in Gdańsk, Faculty of Computer Science and New Technologies, Poland, ORCID: 0000-0001-9436-7142, e-mail: jpiwnik@wsb.gda.pl;*

***Research limitations/implications:*** *The study was limited to enterprises operating in Poland, and its findings may not fully capture international differences. Future research should explore cost–benefit analysis of cybersecurity investments and longitudinal changes in corporate resilience.*
***Practical recommendations:*** *The findings emphasize the need for integrating cybersecurity governance with corporate communication and crisis management. Organizations should treat data protection as a strategic asset and develop transparent disclosure mechanisms to maintain trust after cyber incidents.*
***Originality/value:*** *The paper offers one of the first empirical insights into the intersection of cybersecurity governance and corporate reputation in Polish enterprises, linking technical safeguards with strategic management perspectives.*

## 1. Introduction

In the era of digital transformation, data has become one of the most valuable strategic assets for enterprises. The increasing integration of digital technologies into business operations has enhanced efficiency, innovation, and competitiveness but has simultaneously introduced new risks associated with cybersecurity and data protection.

Cyber incidents, such as data breaches, ransomware attacks, and information leaks, not only disrupt operational continuity but also significantly undermine corporate reputation and stakeholder trust. As organizations become increasingly dependent on digital ecosystems, effective data protection and reputation management have emerged as critical components of sustainable business strategy.

The growing frequency and sophistication of cyber threats have forced enterprises to adopt comprehensive information security management systems. Standards such as ISO/IEC 27001, the General Data Protection Regulation (GDPR), and national cybersecurity frameworks provide guidance for ensuring compliance, transparency, and resilience. However, implementing these frameworks is not limited to technological investment — it requires a cultural and strategic shift that integrates cybersecurity awareness into all levels of corporate governance.

Corporate reputation, traditionally associated with brand image and stakeholder perception, is now deeply interconnected with an organization's digital trustworthiness. Studies indicate that enterprises that effectively manage cybersecurity and communicate transparently during crises are more likely to maintain stakeholder confidence and recover faster from incidents. Conversely, failure to protect data can result in financial losses, regulatory penalties, and long-term damage to brand credibility.

This paper examines how enterprises manage cybersecurity and data protection as elements of reputation governance in the digital economy. The study analyzes organizational practices, adopted standards, and managerial approaches that strengthen resilience and stakeholder confidence.

It contributes to the growing body of knowledge linking cybersecurity governance with corporate reputation management, offering both theoretical insight and practical implications for managers seeking to enhance digital trust and organizational sustainability.

## 2. Literature Review

The enterprise cybersecurity domain has evolved into a multi-layered management problem that integrates technology, governance, legal compliance, and organizational culture. At its core, cybersecurity safeguards the confidentiality, integrity, and availability of information assets through coordinated technical, organizational, and procedural controls (Cieszkowska *et al.,* 2023). Canonical frameworks such as the CIA triad - codified across ISO/IEC 27001 and ISO/IEC 27032 - anchor policy and control design (Antczak, Dębicka, and Nowakowska-Grunt, 2023; Stallings, 2017).

In parallel, zero-trust architectures remove default trust from networks by enforcing continuous authentication and authorization, reflecting contemporary perimeterless environments (Rose, 2020). On the legal plane, GDPR created binding rules for the collection, processing, storage, and timely breach notification of personal data across the EU, making structured privacy and security governance a compliance imperative.

Empirical reporting shows a sharp rise in incident volumes and complexity, with phishing and ransomware dominating attack vectors and often operating in tandem (CERT Polska, 2023; Tetteh, 2024). Sectoral exposure is pronounced in media, retail, courier services, and healthcare, where high-value data and service criticality amplify impact (CERT Polska, 2023; Kubalski *et al.,* 2023).

The commercialization of cybercrime-as-a-service further lowers attacker skill thresholds and scales threat accessibility (Tetteh, 2024). For complex data environments, organizations increasingly pair technical controls with structured methods such as threat modeling (STRIDE) to prioritize mitigations early in the system lifecycle (Kruk, 2021), and apply multi-criteria decision tools to Big Data security posture (Attaallah *et al.,* 2022).

Financially, incident costs are substantial. Average breach costs have reached multi-million-dollar levels globally and within EMEA (Kubalski *et al.,* 2023; Strzelecka,

Szafraniec-Siluta, and Szczepańska-Przekota, 2022). Ransom demands alone can amount to 0.7%–5% of annual revenues, intensifying budgetary trade-offs and post-incident remediation burdens (Kubalski *et al.,* 2023).

High-profile cases (e.g., Yahoo, Target) illustrate the long tail of reputational damage, market value erosion, and prolonged recovery investments (Perera *et al.,* 2022; Kubalski *et al.,* 2023). Despite this, firms' financial disclosures of cyber risk often remain fragmented or incomplete, constraining external risk assessment and benchmarking (Ferens, 2023).

Mature programs blend formal governance with integrated tooling. Adoption of ISO/IEC 27001 and complementary standards correlates with stronger risk management and auditability; the certification has become widespread, with particularly high penetration in transport, forwarding, and logistics (Antczak *et al.,* 2023; Kubalski *et al.,* 2023).

Effective operating models use security steering committees, continuous risk analysis, and defined post-incident playbooks to shorten detection and response cycles (Guasconi *et al.,* 2019). Yet SMEs frequently underestimate cyber risk and under-invest in controls, leaving material digital assets inadequately protected (Guasconi *et al.,* 2019).

On the technology side, layered defenses - endpoint protection, DLP, network monitoring, segmentation - paired with tested backup and recovery (including near-real-time copies and "digital vaults") underpin resilience to data-destruction and extortion attacks (CERT Polska, 2023; Charlinski, 2024).

Nonetheless, legacy infrastructure and integration debt often hinder deployment of modern controls and slow cloud transitions, where Poland lags Western Europe (Antczak *et al.,* 2023; Aftowicz *et al.,* 2020). Case evidence underscores that centralized data governance-harmonized definitions, shared data warehouses, standardized reporting (e.g., XBRL) - reduces reporting errors and improves decision readiness (Dilip and Nandi, 2022), while fragmented data ownership degrades situational awareness during crises (Wong, 2025).

Regulatory trajectories heighten minimum security baselines and accountability. NIS2 mandates defined risk-management frameworks, incident handling, supply-chain security, and oversight, with significant sanctions for non-compliance (Charlinski, 2024). GDPR obligations - such as 72-hour breach notification - tie procedural readiness to legal exposure (Prasołek and Kiełbratowska, 2020).

Internationally, data-localization pressures and diverging governance models (e.g., US–China technology rivalry) complicate cross-border data operations and necessitate adaptable compliance architectures (Tian *et al.,* 2024; Antczak *et al.,* 2023). Heightened hybrid threats expand enterprise risk beyond corporate boundaries into national security domains, reinforcing public–private collaboration needs (Lucini *et al.,* 2022; Ojdana-Kościuszko, 2024).

Supply-chain exposure has strategic salience: heavy reliance on extra-EU inputs (e.g., active pharmaceutical ingredients sourced from China) introduces concentration and integrity risks that cascade into information security and operational continuity (Mierzejewski, Nawrotkiewicz, and Kaczyński, 2025). Recent Polish evidence shows rapid uptake of digital channels, but complex decisions still favor hybrid, trust-anchored models blending online access with human advice (Wojak *et al.,* 2025).

Cyber incidents propagate operational and relational harms: production stoppages, degraded service quality, blocked financial processes, and disrupted client communications (Sakowska-Baryła, 2022). Trust erosion is widespread - attempts to use forged documents were noted by nearly a third of organizations, amplifying authenticity and fraud-prevention requirements (Olszewska *et al.,* 2024).

Because social media accelerates narrative formation, even modest shares of negative content can materially affect brand perception, making crisis communication and real-time monitoring essential components of resilience (Firla *et al.,* 2019; Kobylański, 2024; Sobolewska *et al.,* 2017).

Transparent disclosure, pre-approved scenarios, and regular training of communication teams mitigate reputational spillovers (Kubalski *et al.,* 2023; Kobylański, 2024). Broader digital-trust programs link culture to measurable reductions in exposure (Olszewska *et al.,* 2024).

The literature consistently shows rising incident frequency and cost, effectiveness of standard-based management systems when embedded in governance, and the centrality of recovery readiness and transparent communication to limit long-tail damage. However, several gaps remain. First, SME-specific evidence on the real-world effectiveness and cost–benefit of integrated (technology + governance + culture) programs is limited (Guasconi *et al.,* 2019).

Second, rigorous post-incident disclosure and stakeholder-trust studies are scarce, despite recurring findings on under-reporting and fragmented financial risk communication (Ferens, 2023). Third, the intersection of cloud/AI adoption with zero-trust, data governance, and supply-chain assurance warrants granular evaluation in Polish enterprises that face legacy integration constraints (Antczak *et al.,* 2023; Aftowicz *et al.,* 2020).

Finally, the geopolitical diffusion of cyber risk into operational continuity (transport/energy) calls for models that jointly analyze corporate security posture and sector-level interdependencies (Lucini *et al.,* 2022; Ojdana-Kościuszko, 2024; Mierzejewski *et al.,* 2025).

## 3. Research Methodology and Case Description

Research on strategies and tools supporting cybersecurity management in enterprises aimed to analyze methods of protecting personal data and responding to contemporary digital threats. Its main objective was to identify how organizations approach information security in the context of advancing digitalization and what solutions they use to minimize the risk of cyber incidents.

The research was exploratory in nature and allowed for the identification of existing practices and the assessment of the level of awareness of employees and management in the area of cybersecurity.

A quantitative approach was used to achieve the research objective, which made it possible to obtain a cross-sectional picture of the phenomenon under study. The data was collected using a diagnostic survey method via an online questionnaire. This solution made it possible to reach respondents representing various industries, job levels, and professional experience.

The questionnaire was designed to collect information on the data protection strategies and tools used, the level of awareness of threats, and the preparedness of organizations to respond to cyber incidents.

The study verified hypotheses assuming that companies implement specific cybersecurity management strategies and tools that affect the effectiveness of personal data protection. It also examined whether having a formal strategy and using modern technical tools translates into a higher level of security, and whether regular employee training supports the effectiveness of these activities.

In addition, the relationship between the degree of digitization of a company and the scale of investment in information security management systems was analyzed.

The research process was preceded by a preparatory stage, which included an analysis of the literature on the subject and a review of existing studies on cybersecurity management in enterprises. This allowed us to identify key areas requiring empirical verification and to develop a research tool tailored to the specific nature of the subject matter.

Particular attention was paid to aspects related to the practical implementation of security strategies, the use of protective technologies, and the development of an organizational culture conducive to the secure use of IT systems.

As a result, the study made it possible to obtain data describing the actual activities undertaken in organizations in the context of information protection and cyber risk management.

## 4. Research Results

The survey questionnaire consisted of three parts: introductory, main, and personal details. The first part contained information about the purpose of the study and a guarantee of anonymity, the main part included questions about strategies, tools, and awareness of threats, while the personal details section allowed for the collection of demographic and professional data about the respondents.

The survey results present respondents' answers regarding strategies and tools supporting cybersecurity management in enterprises. The data presented includes information on the solutions used for personal data protection, the level of awareness of threats, and the measures taken to ensure information security.

The results show the state of advancement of cybersecurity measures in the organizations surveyed and provide a basis for further analysis of the individual issues presented in the following sections of this chapter.

The survey results indicate that all companies covered by the analysis have a formal cybersecurity management strategy. Affirmative responses account for 100% of the total, which means there are no negative responses. These data show that each of the organizations surveyed has implemented solutions that formalize their approach to cybersecurity.

***Figure 1.*** *Implementation of a formal cybersecurity management strategy in enterprises.*

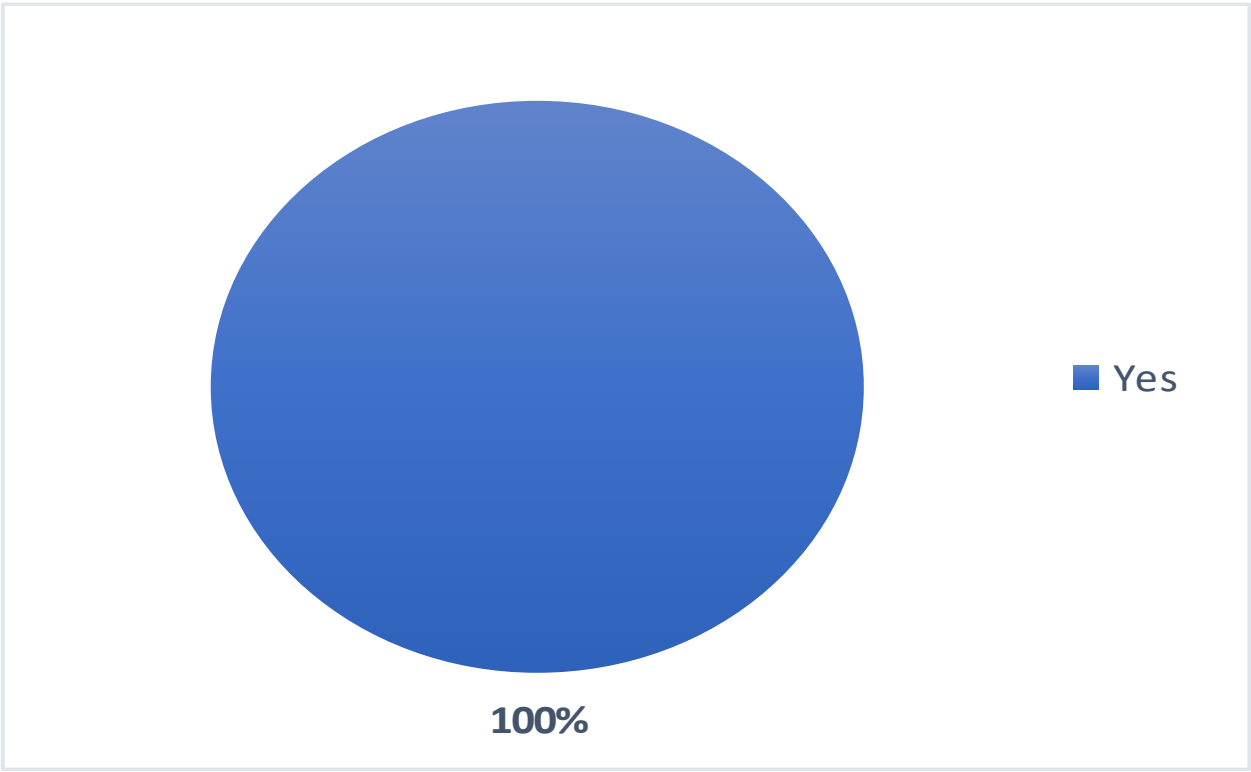

***Source:*** *Own elaboration.*

***Figure 2.*** *Cyber threat awareness assessment.*

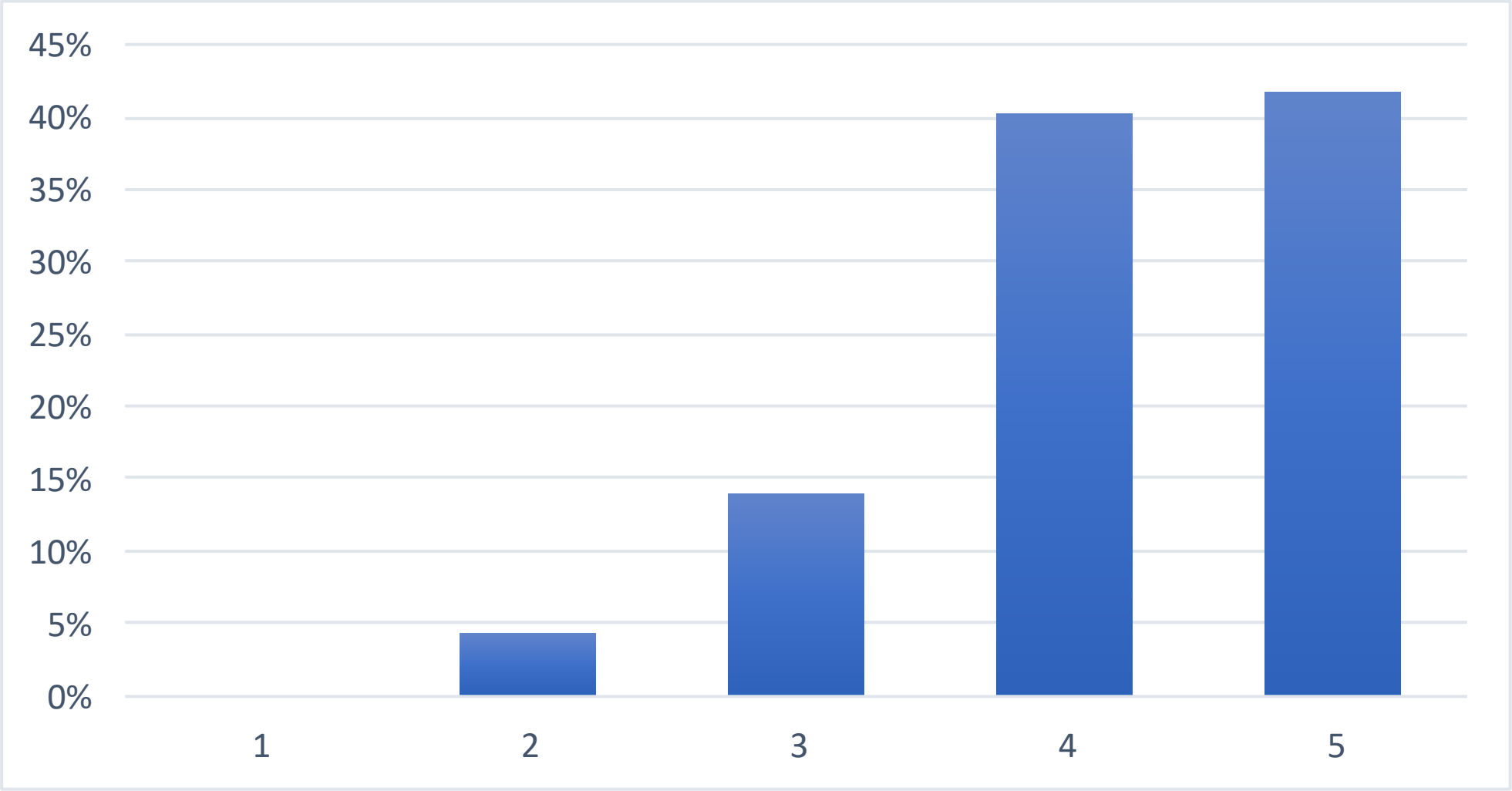

***Source:*** *Own elaboration.*

The chart shows the results of responses to the question: "How would you rate the level of awareness of cyber threats among your company's employees?" The rating scale has five levels, where 1 represents the lowest level of awareness and 5 represents the highest.

The horizontal axis of the chart shows the rating levels, while the vertical axis shows the number of companies that assigned a given value.

The results show a clear predominance of high ratings for the level of awareness of digital threats among employees. Most respondents rated this level as 4 or 5, confirming the prevailing belief that employees have adequate knowledge of how to recognize and avoid potential threats in cyberspace. The small number of lower ratings (1–3) suggests that cases of low awareness in this area are sporadic and affect a limited number of companies.

***Figure 3.*** *Data protection technologies and solutions*

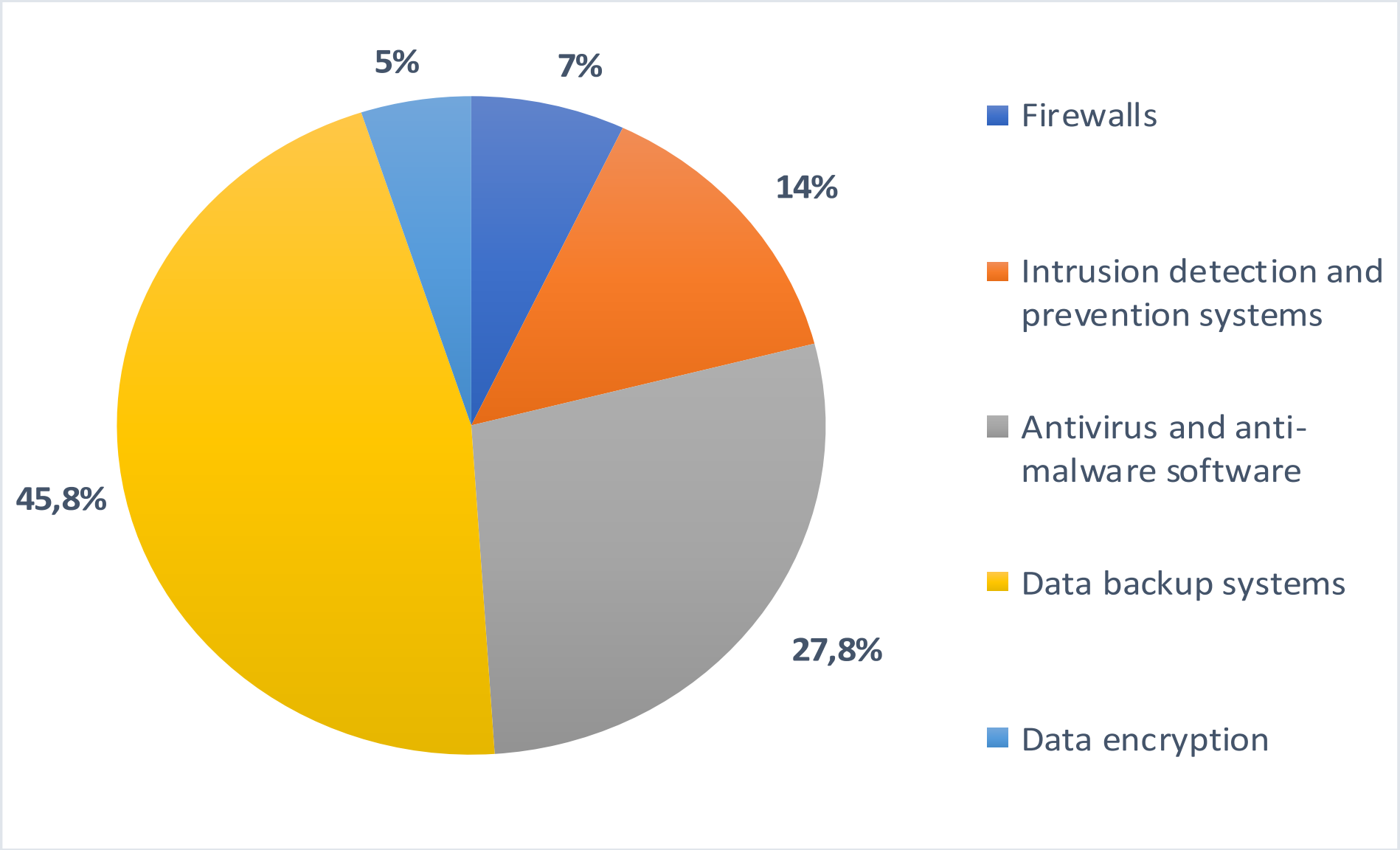

***Source:*** *Own elaboration.*

The pie chart shows the answers to the question: "What are the main technologies and solutions used in your company to protect data?" Each segment represents the percentage of companies indicating a given technology.

The most commonly used solution is data backup systems (45.8%), considered a key element of security strategy. This is followed by antivirus and antimalware software (27.8%) and IDS/IPS intrusion detection and prevention systems (14%).

Firewalls and data encryption account for a smaller share, but despite being less frequently mentioned, they remain important elements of information protection.

The pie chart shows the answers to the question: "Has your company implemented cybersecurity standards and norms, such as ISO/IEC 27001 or ISO/IEC 27032?" The largest share of responses was "Yes" (65.3%), indicating the dominance of organizations that have implemented standards.

The response "I don't know" was chosen by 19.4% of respondents, while 15.3% of companies declared that they had not implemented these standards.

***Figure 4.*** *Implementation of cybersecurity standards and norms*

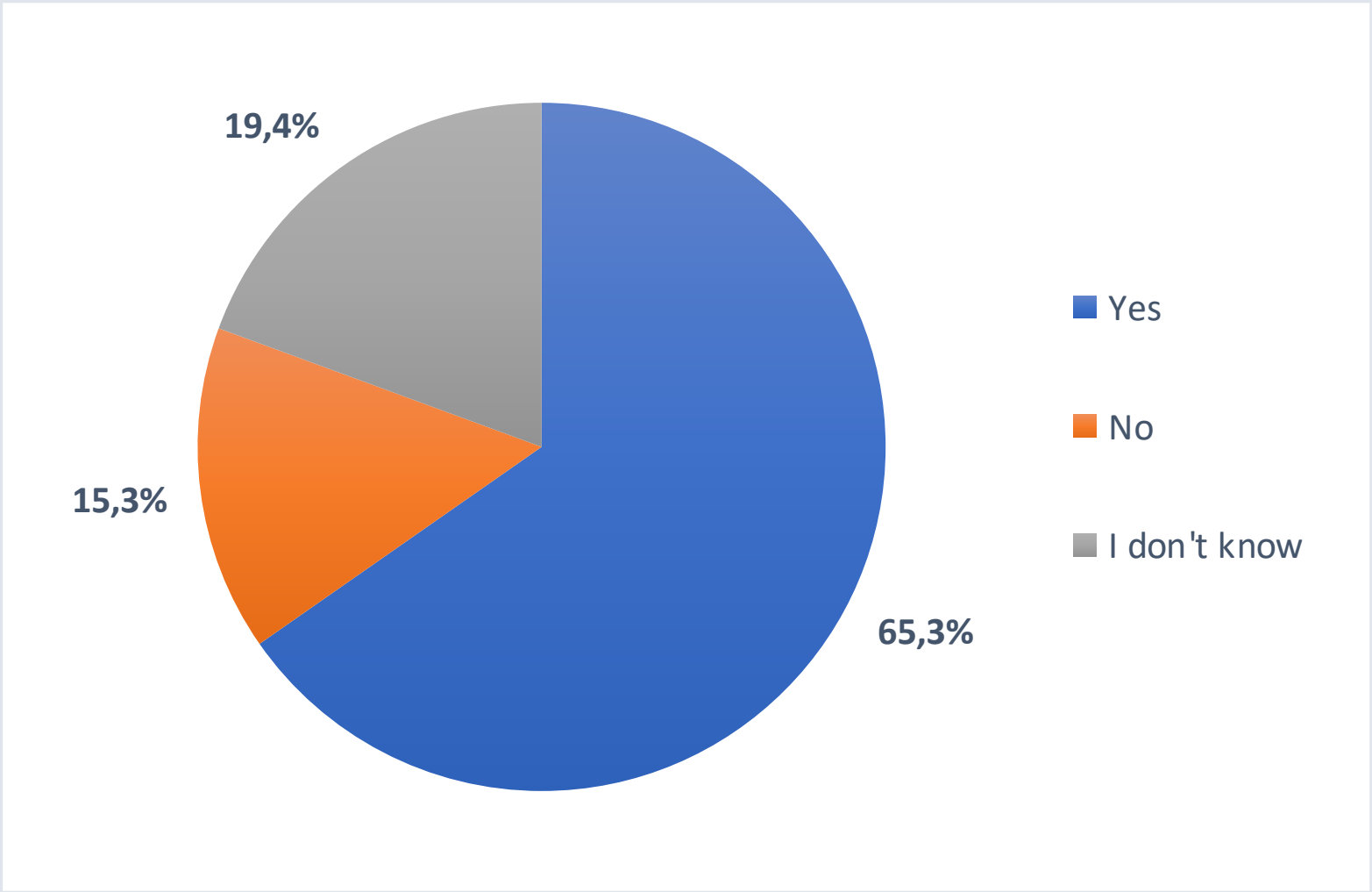

***Source:*** *Own elaboration.*

***Figure 5.*** *Frequency of IT security audits or penetration tests*

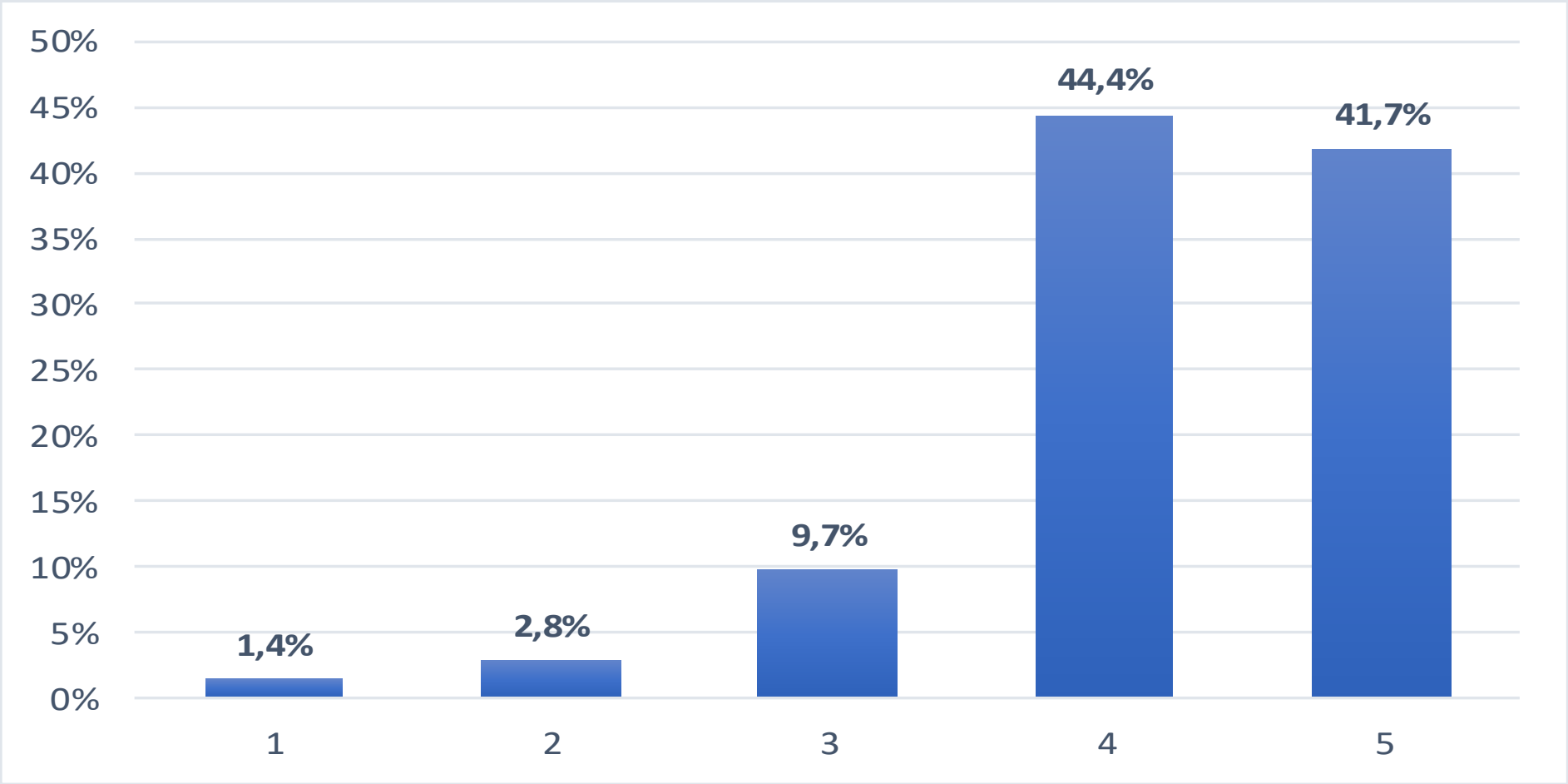

***Source:*** *Own elaboration.*

The chart shows the answers to the question: "How often does your company conduct IT security audits or penetration tests?" The scale from 1 to 5 shows the frequency of these activities, where 1 means the least frequent and 5 means the most frequent audits. The results indicate that most companies regularly perform IT security audits and penetration tests – over 86% of responses fall within the 4 and 5 ranges. The small number of responses at levels 1–3 indicates a limited scale of sporadic control activities in this area.

***Figure 6.*** *Age distribution of respondents*

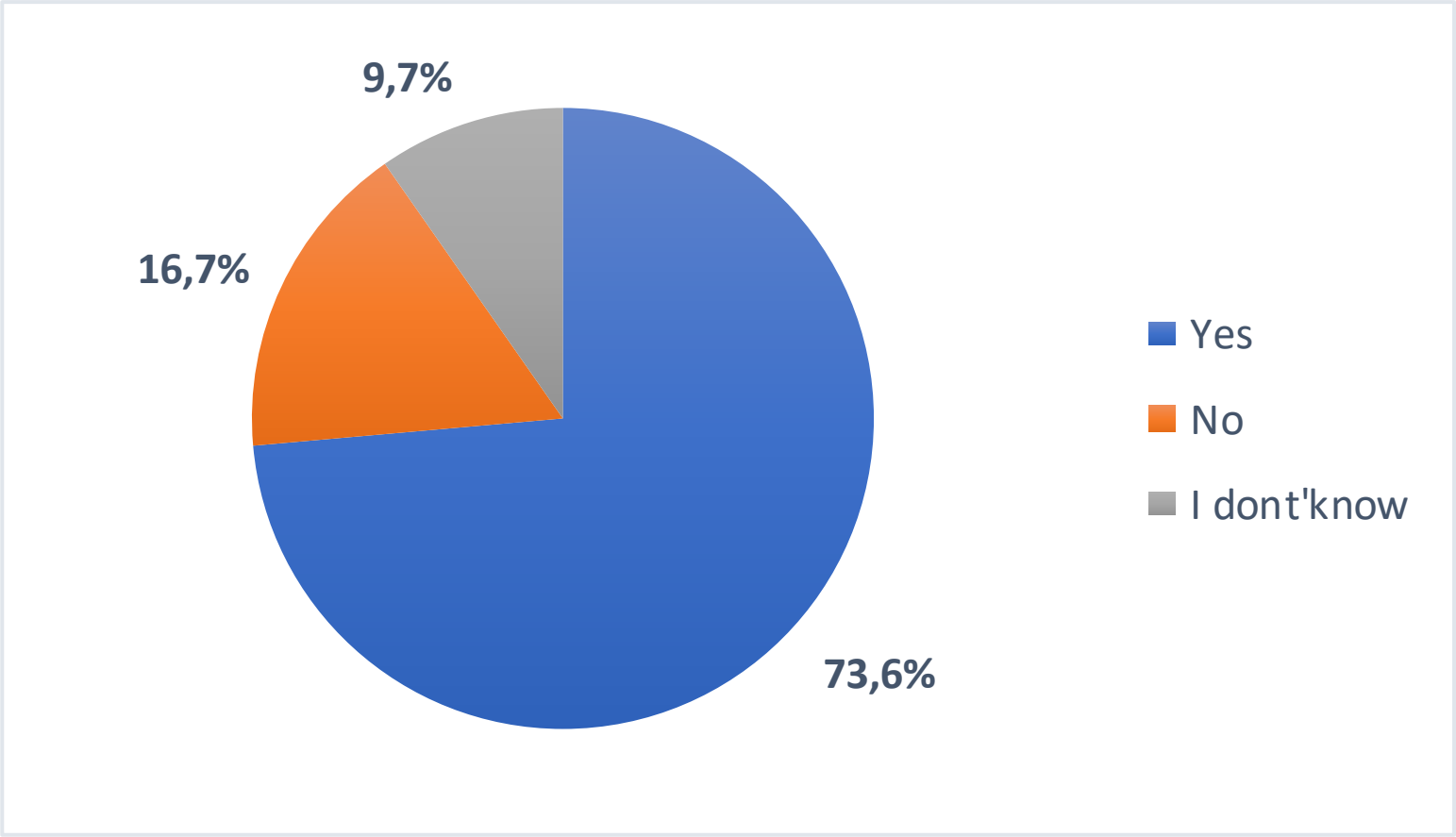

***Source:*** *Own elaboration.*

***Figure 7.*** *Effectiveness of the data protection measures implemented*

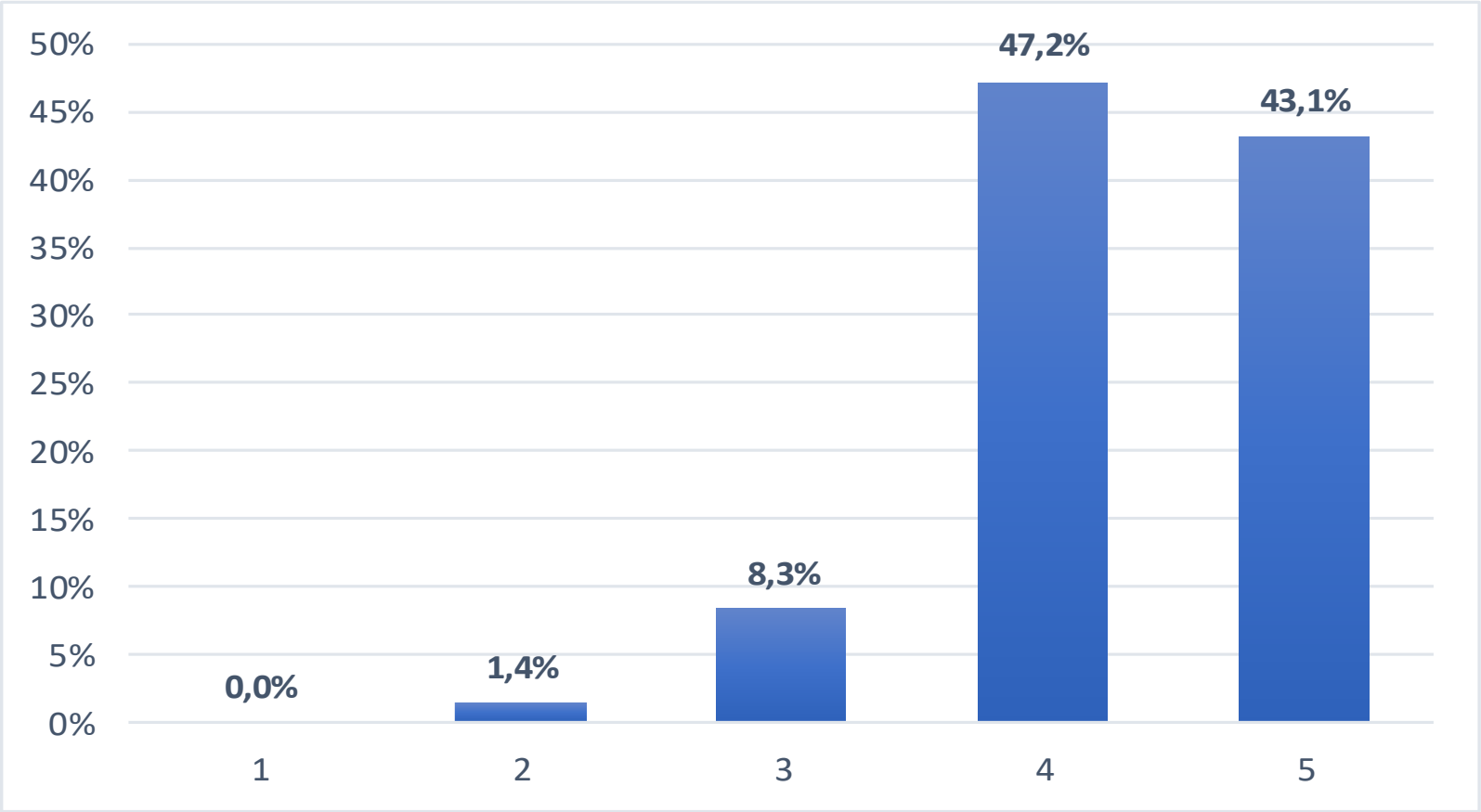

***Source:*** *Own elaboration.*

The pie chart shows the answers to the question: "Does your company have formal procedures for responding to cybersecurity incidents?" Most respondents (73.6%) confirmed the existence of such procedures in their organizations. 16.7% of companies reported a lack of formal rules, while 9.7% of respondents were unaware of their implementation.

The chart shows the answers to the question: "How do you rate the effectiveness of the personal data protection measures implemented in your company?" The rating scale from 1 to 5 covers levels from the lowest to the highest rating of effectiveness. The vast majority of respondents (over 90%) rated the effectiveness of the measures implemented at 4 or 5, which indicates a positive assessment of the actions taken in the field of personal data protection. A small percentage of ratings of 1–3 (approximately 9.7%) indicates a limited number of companies that perceive their solutions as less effective.

***Figure 8.*** *Experience of a cybersecurity incident in the last 12 months*

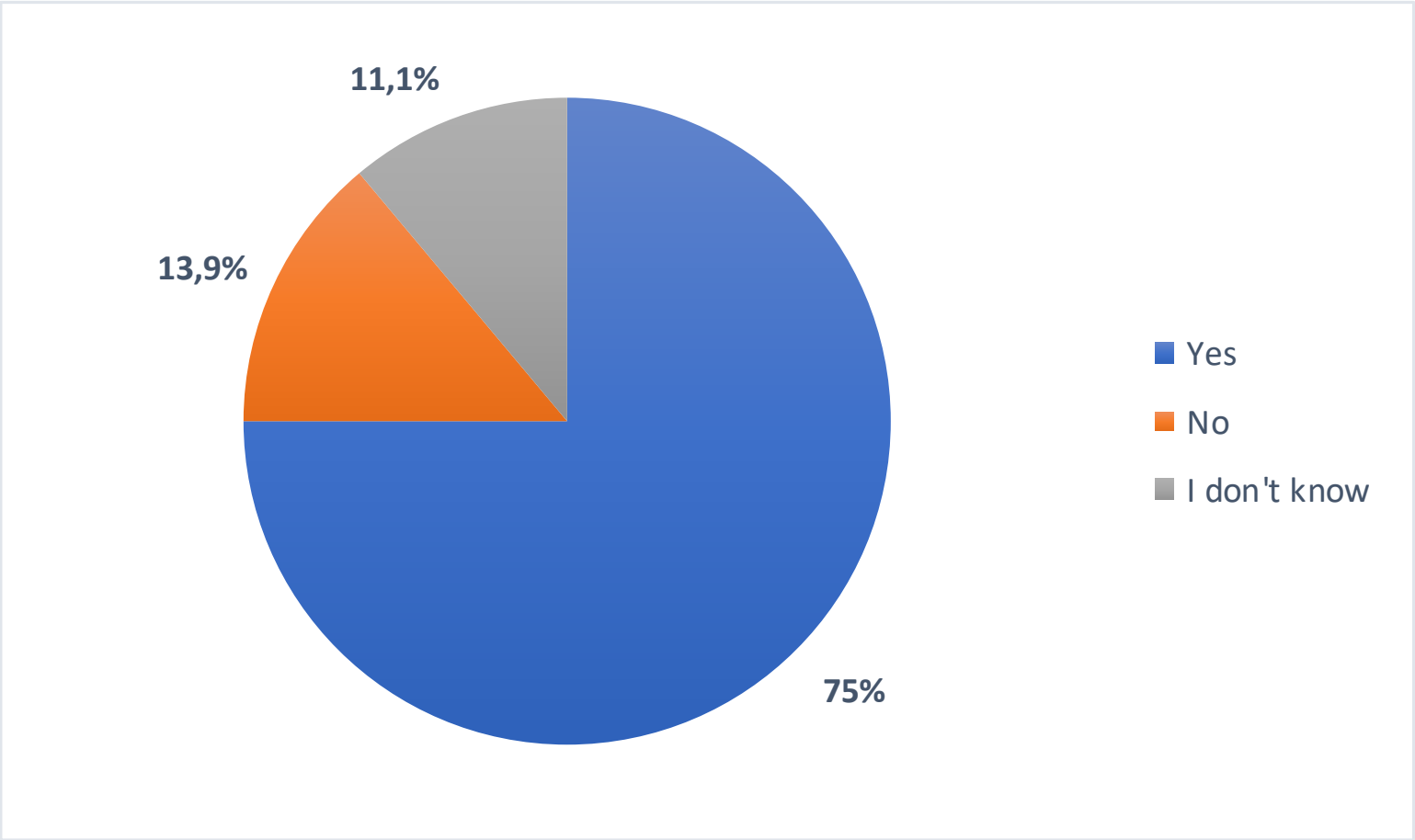

***Source:*** *Own elaboration.*

The pie chart shows the answers to the question: "Has your company experienced a cybersecurity incident (e.g., data breach, ransomware attack, phishing) in the last 12 months?" Most companies (75%) confirmed that such an incident had occurred. 13.9% of respondents reported no incidents, while 11.1% did not know whether such an incident had occurred in their organization.

The pie chart shows the answers to the question: "If your company has experienced a cybersecurity incident, what were the main consequences of that incident?" The data only applies to companies that confirmed an incident in the last 12 months.

The most frequently cited consequence was loss of reputation or customer trust (36.1%). This was followed by business downtime (29.2%) and financial losses and legal or regulatory consequences (12.5% each). The least frequently mentioned consequence was data loss (9.7%), which, despite the lower percentage, remains a significant effect of cybersecurity incidents.

***Figure 9.*** *Consequences of a cybersecurity incident for businesses*

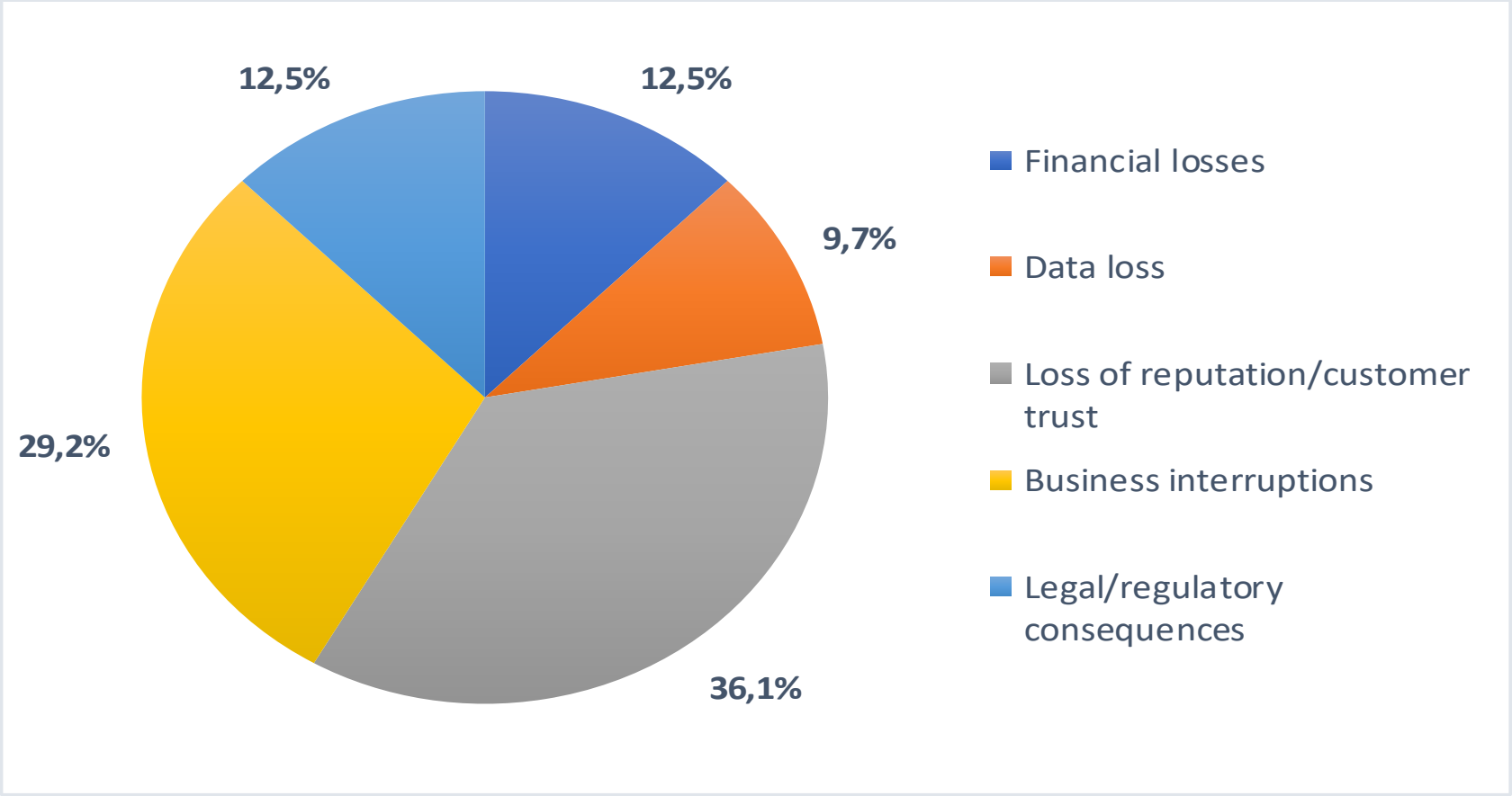

***Source:*** *Own elaboration.*

***Figure 10.*** *Monitoring online reputation in the context of data security and potential threats*

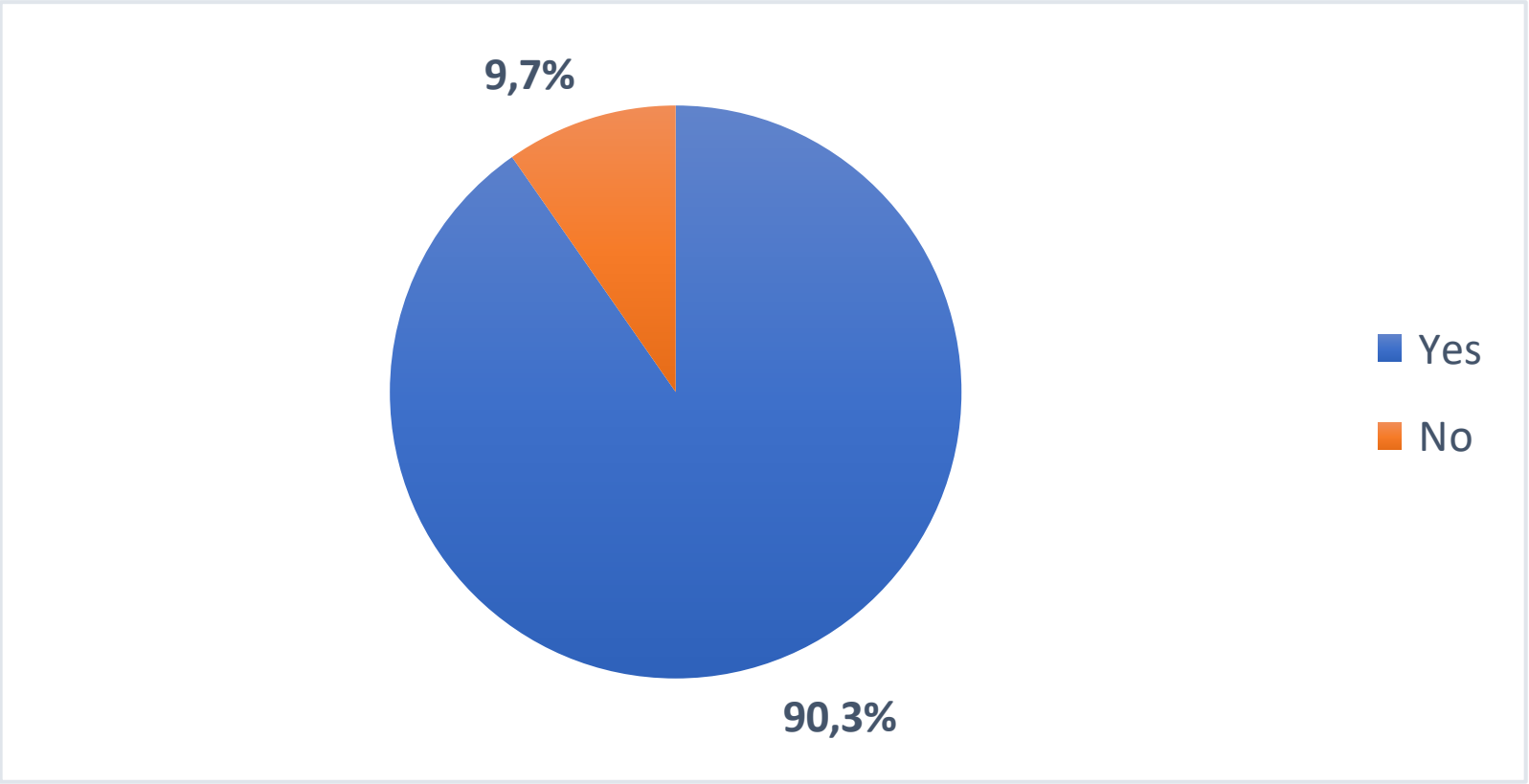

***Source:*** *Own elaboration.*

The pie chart shows the answers to the question: "Does your company monitor its reputation on the Internet in the context of data security and potential cyber threats?" The vast majority of respondents (90.3%) confirmed that they conduct such monitoring, while 9.7% of companies declared that they do not. The results indicate that online reputation monitoring is a common practice among the organizations surveyed.

***Figure 11.*** *Investment in cybersecurity as a cost/investment*

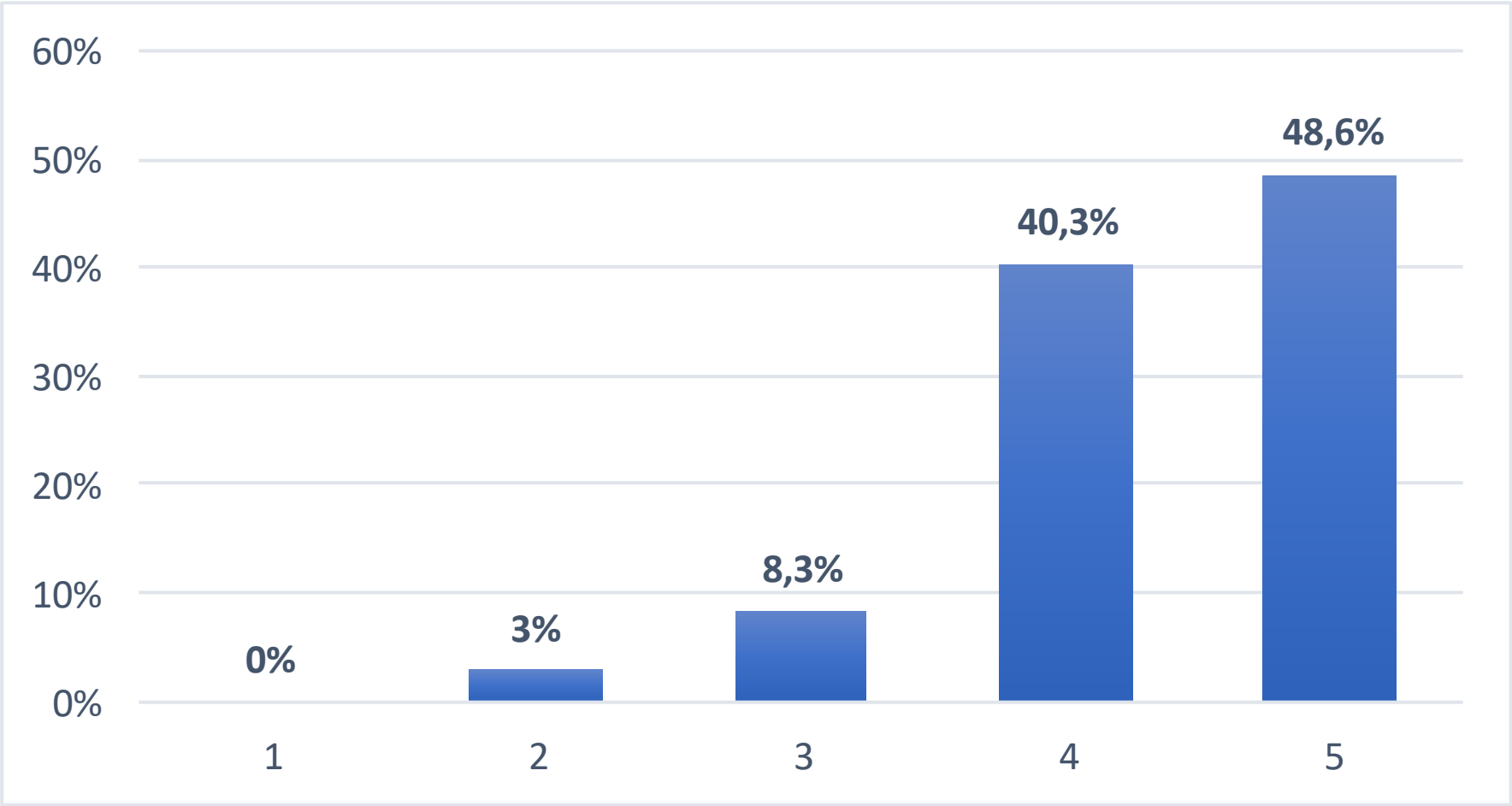

***Source:*** *Own elaboration*

The chart shows the answers to the question: "In your opinion, are investments in cybersecurity treated as a cost or as an investment in the future and stability of the company?" The scale from 1 to 5 reflects the degree of conviction, where 1 means perceiving the expenditure as a cost, and 5 as an investment in the development and security of the organization.

The majority of respondents (over 88%) considered cybersecurity spending to be an investment in the future and stability of the company. A small percentage of lower ratings indicates that only a few organizations treat these activities solely as an operating cost.

The pie chart shows the distribution of responses regarding the respondents' place of residence. Most of the respondents (75%) live in cities with up to 150,000 inhabitants, while 15.3% live in cities with more than that number. A small proportion of respondents live in towns with up to 50,000 inhabitants, and the smallest percentage are people living in rural areas.

The pie chart shows the structure of the industries from which the respondents come. The largest group is represented by the service industry (30.6%), followed by trade (26.4%). 12.5% of respondents represent the healthcare and manufacturing sectors. The IT industry accounts for 9.7% of survey participants, while the smallest share is represented by the financial sector and local and public administration.

***Figure 12.*** *Respondents' place of residence*

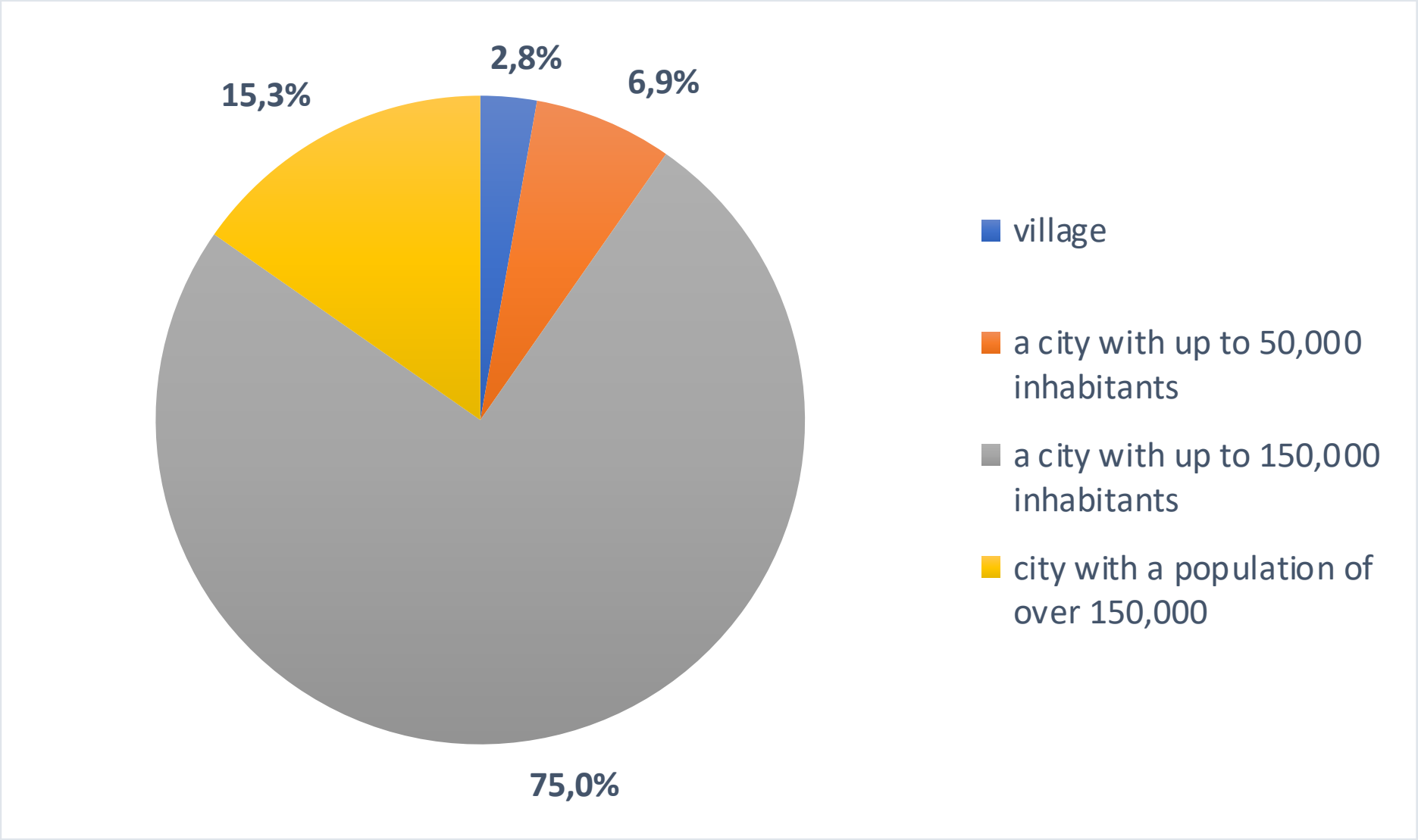

***Source:*** *Own elaboration.*

***Figure 13.*** *Respondents' industry*

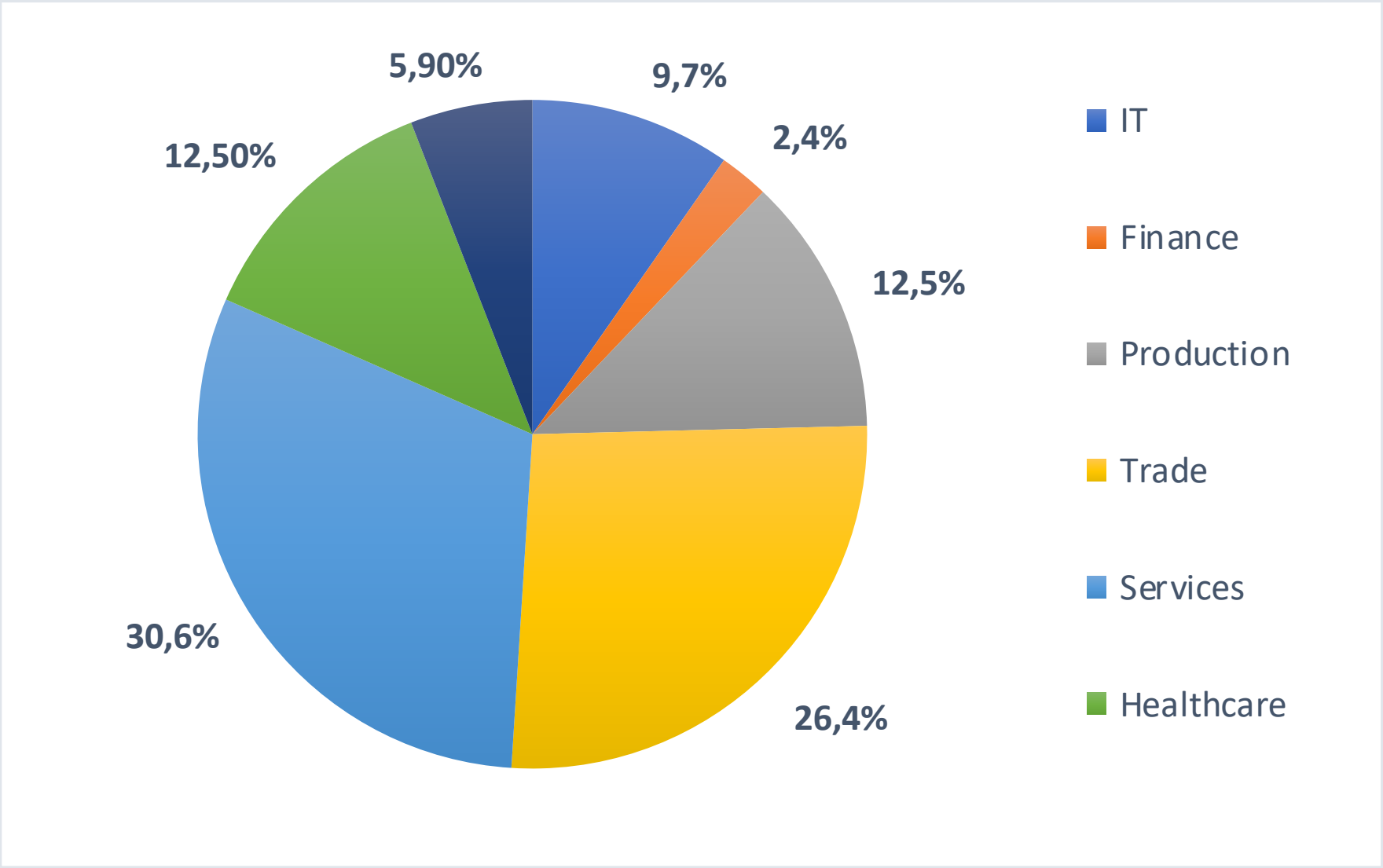

***Source:*** *Own elaboration.*

***Figure 14.*** *Size of the enterprise*

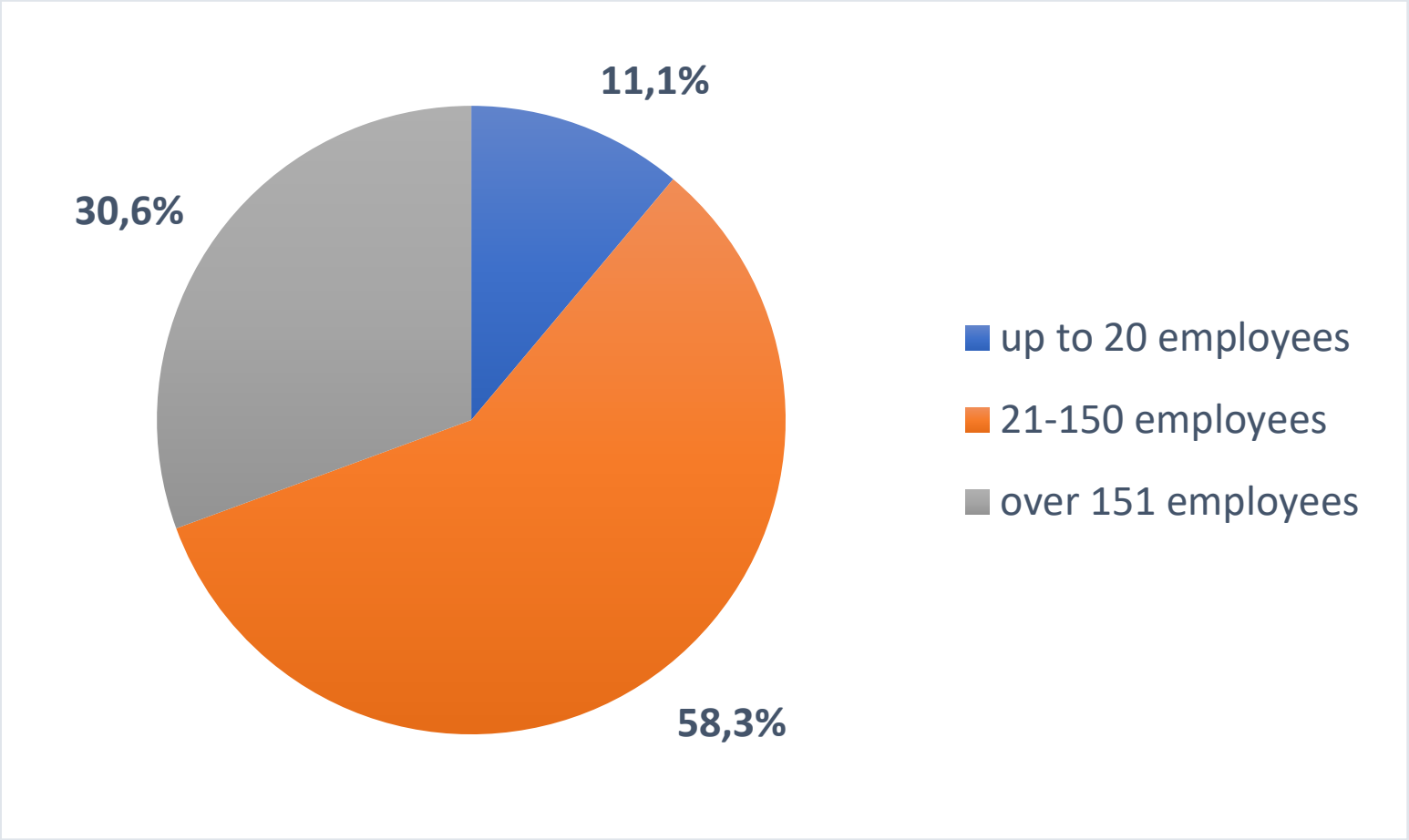

***Source:*** *Own elaboration.*

The pie chart shows the distribution of responses regarding the size of the companies where respondents work. The largest group consists of representatives of medium-sized companies employing between 21 and 150 people (58.3%). The next group consists of people employed in large companies with more than 151 employees (30.6%), while the smallest percentage of respondents (11.1%) come from small companies employing up to 20 people.

The research provided information on the state of cybersecurity management in enterprises and the strategies and tools used in this area. The results show that most of the organizations surveyed have formal strategies and procedures related to information security, as well as implemented standards such as ISO/IEC 27001 and ISO/IEC 27032. The most commonly used technical solutions included data backup systems, antivirus and antimalware software, as well as intrusion detection and prevention systems. The respondents represented a variety of industries, including services, trade, manufacturing, healthcare, and IT, which provided a cross-sectional view of the situation in different areas of the economy. Most of the survey participants came from medium-sized companies with 21 to 150 employees, as well as companies operating in cities with a population of up to 150,000.

## 5. Conclusions and Future Research Implications

The analysis of cybersecurity management practices among Polish enterprises reveals a high degree of formalization and technological maturity in data protection and risk management frameworks. All organizations surveyed reported having formal cybersecurity strategies, reflecting broad compliance with both regulatory and governance requirements. The adoption of international standards such as ISO/IEC 27001 and ISO/IEC 27032 was widespread (65.3%), and regular IT security audits or penetration tests were confirmed by more than 86% of respondents, suggesting institutionalized oversight and continuous improvement processes. These practices align with global recommendations emphasizing governance, procedural readiness, and integration of technological safeguards into enterprise management systems.

The findings indicate that employee awareness of cyber threats is consistently high, with most organizations rating awareness at levels 4–5 on a five-point scale. This outcome underscores the success of awareness programs and internal communication mechanisms in promoting security-conscious behavior.

Technical controls were dominated by data backup systems (45.8%), antivirus and antimalware tools (27.8%), and intrusion detection and prevention systems (14%). Together, these measures demonstrate that organizations prioritize operational continuity and data resilience as central components of their cybersecurity posture. The research also found that 73.6% of enterprises maintain formal incident response procedures, confirming procedural readiness to manage and mitigate cyber incidents effectively.

Despite the high level of preparedness, 75% of companies experienced a cybersecurity incident in the previous twelve months, indicating that formal structures and tools, while essential, do not fully eliminate exposure. The most frequently reported consequence of incidents was reputational damage and loss of customer trust (36.1%), followed by operational disruptions (29.2%) and financial or regulatory impacts (12.5% each).

These results highlight the reputational dimension of cybersecurity and the importance of integrating crisis communication and online reputation monitoring into risk management. Indeed, 90.3% of surveyed organizations actively monitor their online presence for signs of potential data-security threats, demonstrating growing awareness of digital trust as a strategic asset.

The study's results further reveal a shift in managerial perception of cybersecurity from a necessary cost toward a forward-looking investment. Over 88% of respondents considered cybersecurity spending as an investment in long-term stability and organizational resilience.

This aligns with prior literature emphasizing that cybersecurity governance, when embedded within organizational culture and supported by continuous employee training, enhances both operational reliability and stakeholder confidence. The research thereby reinforces the position that cybersecurity is not solely a technical or compliance issue, but a multidimensional management process interlinking technology, human factors, and strategic foresight. These findings corroborate prior studies (Antczak *et al.,* 2023; Kubalski *et al.,* 2023; Charlinski, 2024) that stress the importance of standardized frameworks and proactive governance in reducing cyber risk exposure. They also extend earlier analyses by illustrating the specific maturity level of Polish enterprises and the convergence between regulatory compliance (GDPR, NIS2) and managerial perception of cybersecurity as a value-creating function. However, limitations persist: the study's sample, while diverse in industry representation, focused on a national context and did not assess the cost–benefit relationship of integrated security programs or long-term post-incident recovery outcomes.

Future research could expand this analysis through longitudinal studies exploring how organizational security cultures evolve over time, especially under conditions of accelerated digitalization and hybrid work. Comparative cross-sector analyses would also help identify industry-specific barriers to the implementation of advanced security frameworks and technologies. Additionally, exploring the intersection of AI, zero-trust architectures, and supply-chain security would provide valuable insights into how enterprises can adapt governance models to an increasingly complex and interdependent threat environment.

**References:**

Aftowicz, A., Bartniak, W., Bilski, P., Borkowska, I., Chojnacka, P., Chrońska, M. 2020. Trendy w biznesie (Tom III). SIZ Wydawnictwo. https://www.wydawnictwo-siz.pl/wp-content/uploads/2020/10/Trendy-w-biznesie-DRUK.pdf#page=17.

Antczak, J., Dębicka, E., Nowakowska-Grunt, J. 2023. Wybrane aspekty zarządzania bezpieczeństwem informacji w organizacjach w świetle współczesnych wyzwań gospodarki. Przykład przedsiębiorstw działających w Polsce. Studia Wschodnioeuropejskie, 2(19), 311-335.

Attaallah, A. 2022. Analyzing the big data security through a unified decision-making approach. Intelligent Automation & Soft Computing, 32(2), 1072-1088.

CERT Polska. 2023. Raport o stanie bezpieczeństwa cyberprzestrzeni Rzeczypospolitej Polskiej w 2022 roku. CSIRT GOV.

Charlinski, B. 2024. Zgodność z NIS2 – cyberbezpieczeństwo w zgodzie z dyrektywą europejską. Dell Technologies.

Cieszkowska, E. 2023. Sektorowa Rama Kwalifikacji dla Cyberbezpieczeństwa. Instytut Badań Edukacyjnych.

Dilip, A., Nandi, D. 2022. From Data Management to Data Governance: Experience of the Reserve Bank of India. Reserve Bank of India. https://www.isi2023.org/media/abstracts/ottawa2023_1ea38b24cac91ae47c8c4cea564f650b.pdf.

Ferens, A. 2023. Ujawnienie szczególnych obszarów ryzyka w wybranych sprawozdaniach w dobie kryzysu. Zeszyty Naukowe Wydziału Zarządzania GWSH, 20, 23-32.

Firla, K., Czerwiński, J., Zielińska, A.M., Szczepkowski, J., Wąchol, J., Karczewska, O. 2019. Współczesne wyzwania cyfryzacji – przegląd i badania. Wydawnictwo Naukowe TYGIEL. https://bc.wydawnictwotygiel.pl/public/assets/357/Wspolczesne%20wyzwania%20cyfryzacji%20-%20przeglad%20i%20badania.pdf#page=80.

Guasconi, F., Papadopoulou, G., Sharkov, G., Bulavrishvili, D., Oteiza, S., Berens, H. 2019. Wdrażanie ISO/IEC 27001. Instytut Kościuszki.

Kruk, T. 2021. Praktyczne modelowanie zagrożeń dla systemów teleinformatycznych z wykorzystaniem modelu STRIDE. Pomiary Automatyka Robotyka, 25(4), 93-97.

Kubalski, J., Pasztaleniec, M., Krawczyński, K., Jurek, P., Kraśniewska, K., Zięcina, D., Paszek, K., Mazur, M. 2023. Cyberbezpieczeństwo – wyzwania dla biznesu. Dagma Bezpieczeństwo IT.

Kobylański, D. 2024. Co wypada, a co nie wypada mówić, czyli pauperyzacja języka polskiej polityki. Uniwersytet Marii Curie-Skłodowskiej w Lublinie. https://www.archaegraph.pl/lib/l231bv/Jezyk-w-mediach-1-m2j7a6ow.pdf#page=95.

Lucini, B., Yilmaz, K., Atamuradova, F., Karpiuk, M., Kanayama, R.D., Barone, D.M. 2022. Sicurezza, terrorismo e società, 16, 1-123. https://www.sicurezzaterrorismosocieta.it/wp-content/uploads/2022/12/SicTerSoc16.pdf#page=114.

Mierzejewski, D., Nawrotkiewicz, J., Kaczyński, P.M. 2025. Jaka polityka UE wobec Chin? Fundacja Kazimierza Pułaskiego.

Ojdana-Kościuszko, M. 2024. Ewolucja i wyzwania polskiego systemu cyberbezpieczeństwa. Securitate et Defensione, 1(10), 128-146.

Olszewska, A., Gajak-Skwira, A., Podpłoński, R., Kopniak, M., Szczepański, K., Trętowski, R. 2024. Zaufanie w cyfrowym świecie. Kierunek: wiarygodność. KIR. https://www.elektronicznypodpis.pl/storage/file/core_files/2025/1/29/26753d16abf9697309853fac533f4371/raport_zaufanie_w_cyfrowym_swiecie_kierunek_wiarygodnosc.pdf.

Perera, S. 2022. Factors affecting reputational damage to organisations due to cyberattack. Informatics, 9(28), 1-25.

Prasołek, Ł., Kiełbratowska, A. 2020. Praca zdalna w praktyce: Zagadnienia prawa pracy i RODO. C.H. Beck.

Rose, S. 2020. Zero Trust Architecture. NIST Special Publication 800-207.

Sakowska-Baryła, M. 2022. Uprawnienia pracodawców w czasie pandemii – praca zdalna w sektorze medycznym. Monitor Prawa Pracy, 2/2022, 29-30. https://www.ksiegarnia.beck.pl/media/product_custom_files/2/0/20959-monitor-prawa-pracy-nr-2-2022-fragment.pdf.

Sobolewska, O., Waszkiewicz, M., Wallis, A., Kempa, E., Rybak, A., Dzikowski, P. 2017. Gospodarka cyfrowa 2016. Zarządzanie, innowacje, społeczeństwo i technologii. Wydział Zarządzania.

Stallings, W. 2017. Computer Security: Principles and Practice. Pearson Education.

Strzelecka, A., Szafraniec-Siluta, E., Szczepańska-Przekota, A. 2022. Ryzyko w działalności przedsiębiorstw – wybrane aspekty. Politechnika Koszalińska.

Tian, Z., Zakaria, A., Latif, A., Prakesh, S., Price, L., Wei, O. 2024. Bridging digital divides: Navigating data governance and security in the U.S.-China technological arena. Fudan-Harvard China-U.S. Young Leaders Dialogue. https://iis.fudan.edu.cn/_upload/article/files/c7/ee/602f356d4b659f81eb518bd2a26b/b204a4fa-83fe-44c9-8c04-41ad4c82717f.pdf.

Tetteh, A.K. 2024. Cybersecurity needs for SMEs. Issues in Information Systems, 25(1), 235.

Wojak, G., Górka, E., Ćwiąkała, M., Baran, D., Świniarski, R., Olszyńska, K., Mrzygłód, P., Frasunkiewicz, M., Ręczajski, P., Zawadzki, D., Piwnik, J. 2025. The Evolution of Insurance Purchasing Behavior: An Empirical Study on the Adoption of Online Channels in Poland. Scientific Papers of Silesian University of Technology. Organization and Management Series, No. 228.

Wong, C. 2025. Data Governance Avenues for Military Cultural Transformation. Digital Policy Hub.